\documentclass[a4paper,11pt]{article}
\pdfoutput=1
\usepackage{jcappub}

\usepackage{tabularx,booktabs}
\newcolumntype{Y}{>{\centering\arraybackslash}X}

\usepackage[utf8]{inputenc}
\usepackage[font=small,labelfont=bf,tableposition=top]{caption}
\usepackage{float}
\usepackage{adjustbox}
\usepackage{arydshln}

\usepackage{amsmath,amssymb,amsbsy,amstext, amsthm, amsfonts, bm}
\usepackage{hyperref}
\usepackage{graphicx}
\usepackage{siunitx}
\usepackage{upgreek}
\usepackage{framed}
\usepackage{wrapfig}
\usepackage{multirow}
\usepackage{subfig}
\usepackage{bbm}
\usepackage[svgnames,dvipsnames,x11names]{xcolor}
\usepackage{enumitem}
\usepackage{fontawesome}

\usepackage[top=2.5cm,bottom=3.cm,right=2.5cm,left=2.5cm]{geometry}

\renewcommand{\d}{\mathrm{d}}

\def\<{\left\langle}
\def\>{\right\rangle}

\def\bx{{\bf x}}

\newcommand{\adsurl}[1]{\href{#1}{ADS}}

\newcommand\ba{\begin{eqnarray}}
\newcommand\ea{\end{eqnarray}}
\newcommand\be{\begin{equation}}
\newcommand\ee{\end{equation}}

\newcommand\gsim{ \lower .75ex \hbox{$\sim$} \llap{\raise .27ex \hbox{$>$}} }
\newcommand\lsim{ \lower .75ex \hbox{$\sim$} \llap{\raise .27ex \hbox{$<$}} }

\newcommand{\vx}{{\mathbf{x}}}

\newcommand{\vk}{{\mathbf{k}}}

\newcommand{\eq}[1]{(\ref{eq:#1})} 
\newcommand{\eqq}[1]{Eq.~(\ref{eq:#1})} 
\newcommand{\fig}[1]{Fig.~\ref{fig:#1}}

\renewcommand{\vec}[1]{\mathbf{#1}}

\newcommand{\vz}{\vec{z}}
\newcommand{\vq}{\vec{q}}

\definecolor{darkgreen}{RGB}{0,120,0}

\title{Galaxy Skew-Spectra in Redshift-Space}

\author[a]{Marcel Schmittfull}
\author[b]{and Azadeh Moradinezhad Dizgah}
\affiliation[a]{School of Natural Sciences, Institute for Advanced Study, 1 Einstein Drive, Princeton, NJ 08540, USA}
\affiliation[b]{ D\'epartement de Physique Th\'eorique,
Universit\'e de Gen\`eve, 24 quai Ernest Ansermet, \\ 1211 Gen\`eva 4, Switzerland}

\emailAdd{mschmittfull@gmail.com, Azadeh.MoradinezhadDizgah@unige.ch}

\abstract{Modern galaxy surveys focus on the galaxy power spectrum or 2-point correlation function to test and constrain cosmological models. 
Additional information comes from higher-order N-point functions, but their analysis is challenging. A simple solution is to compute the cross-power spectrum between the squared galaxy density and the galaxy density. Being simple to measure and to plot, this skew-spectrum shares many of the familiar useful properties of the standard galaxy power spectrum. We show that by computing multiple quadratic fields and correlating them with the density, all contributions to the tree-level redshift-space galaxy bispectrum can be captured with skew-spectra. Using synthetic datasets, we show that our measurement pipeline matches analytical predictions, and that their dependence on galaxy bias parameters and the logarithmic growth rate is as expected theoretically. 
}

\begin{document}
\maketitle
\flushbottom

\section{Introduction}

Galaxy redshift surveys are becoming an increasingly important cosmological probe, which is reflected by the large number of funded and proposed experiments in the coming years, including DESI \cite{Aghamousa:2016zmz}, HSC \cite{HSC}, Euclid \cite{Amendola:2016saw}, Vera Rubin Obervatory/LSST \cite{Abell:2009aa}, SPHEREx \cite{Dore:2014cca}, Roman Telescope/WFIRST \cite{2019arXiv190205569A}, and others. The key summary statistic used to analyze these experiments is the power spectrum or 2-point correlation function. Thanks to extensive work by the community over the past decades, this analysis approach is relatively mature now
(though improvements are still possible, e.g.~related to theoretical errors \cite{Baldauf:2016sjb,Chudaykin:2020hbf}, covariance estimation \cite{Philcox:2020zyp}, wide-angle effects \cite{Yoo:2013zga,Castorina:2019hyr,Castorina:2018nlb}, or optimal weights \cite{Tegmark:1996bz,Ruggeri:2016mac,Pearson:2016jzc,Mueller:2017pop,Castorina:2019wmr}). A less mature analysis approach involves higher-order statistics, like the galaxy 3-point correlation function, or its Fourier-space equivalent, the bispectrum, which adds information on cosmological and nuisance parameters. While different approaches have been proposed to analyze these, including \cite{Scoccimarro:2015bla,Slepian:2015qwa,Slepian:2015qza,Fergusson:2010ia,Schmittfull:2012hq,Gil-Marin:2014sta}, they remain challenging for future surveys that include tens of millions of galaxy redshifts (some challenges include accurate and fast modeling of the signal, accurate modeling of the covariance, inclusion of the window survey function, and computational speed).
To ameliorate some of these issues, it would be useful to have simpler and faster analysis frameworks. For this reason, several proxy statistics for the bispectrum have been introduced in the literature \cite{Regan:2011zq, Obreschkow:2012yb, Schmittfull:2014tca,Chiang:2015pwa}. 
\vskip 4pt

Motivated by this, we investigate galaxy skew-spectra in this paper. These are cross-power spectra between the squared galaxy density and the galaxy density.
With appropriate filters, they can be derived as optimal bispectrum estimators in the limit of weak non-Gaussianity \cite{Schmittfull:2014tca}. Indeed, these skew-spectra have the same Fisher information content for galaxy bias parameters, the amplitude of scalar fluctuations, the amplitude of the primordial bispectrum ($f_\mathrm{NL}$), and the growth factor as the full bispectrum, therefore representing a lossless compression \cite{MoradinezhadDizgah:2019xun}. In our opinion, the main advantage of these skew-spectra is their simple interpretation -- they are functions of a single wavenumber and do not involve any triangles. Potential additional advantages are simplifications when computing covariances and computational speed. In terms of the computational cost, capturing the full information of the bispectrum using the skew spectra requires $\mathcal{O}( N\log N$) operations, where $N=(k_\mathrm{max}/\Delta k)^3$ is the number of 3D Fourier-space grid points at which the fields are evaluated given a small-scale cutoff of $k_\mathrm{max}$. In contrast, accounting for all bispectrum triangles, requires $\mathcal{O}(N^2)$ operations.
\vskip 4pt

As a first step, previous studies investigated these galaxy skew spectra without taking into account redshift-space distortions  \cite{Schmittfull:2014tca,MoradinezhadDizgah:2019xun}. 
These are caused by the fact that observed galaxy redshifts are shifted from their true position by their velocity along the line of sight. Breaking statistical isotropy, they change the structural form of the galaxy bispectrum. As a consequence, the skew-spectra that are optimal for galaxies in real space are not sufficient to extract the full bispectrum information of galaxies in redshift-space. In this paper, we go one step further and ask what skew-spectra are needed to optimally estimate the bispectrum of galaxies in redshift-space. This is a crucial step to make skew-spectra applicable to real data from galaxy redshift surveys. As a first step in that direction, Ref.~\cite{Dai:2020adm} recently considered the redshift-space skew spectrum corresponding to the scale-independent bispectrum monopole and that following from local primordial non-Gaussianity. Here we derive the complete set of skew spectra capturing the information of all of the bispectrum contributions at tree level.
\vskip 4pt

The paper is organized as follows. In Section \S \ref{sec:skew_rsd} we derive the full set of skew spectra corresponding to tree-level redshift-space galaxy bispectrum, accounting for local-in-matter and tidal biases. In Section \S \ref{sec:sims} we compare the theoretical predictions of the skew spectra against their measurements on N-body simulations. Finally, in section \S \ref{sec:con} we draw our conclusions.
\vskip 4pt

Throughout the paper, we use the Fourier convention
\begin{equation}
  \label{eq:6}
  f(\vk) = \int \d^3\vx e^{-i\vk\cdot\vx} f(\vx),\qquad
  f(\vx) = \int \frac{\d^3\vk}{(2\pi)^3} e^{i\vk\cdot\vx} f(\vk).
\end{equation}
Therefore $[\mathbf{\nabla}_\vx f](\vk)=i\vk f(\vk)$.
\vskip 4pt

\section{Skew-spectra from the Tree-level Galaxy Bispectrum in Redshift-space}\label{sec:skew_rsd}

We start by reviewing the leading-order, tree-level perturbation theory model for the bispectrum of galaxies in redshift-space. As we will see, all contributions to the tree-level bispectrum are product-separable in wavevectors and, as a result, their amplitudes can be measured using skew-spectra. At the end of the section, we will derive these skew-spectra following from the tree-level galaxy bispectrum in redshift-space.
\vskip 4pt

\subsection{Galaxy Density in Redshift-space}

Assuming a deterministic relation between galaxy and dark matter density fields (and neglecting higher derivative operators), up to second-order in the matter density field, the galaxy overdensity can be expanded in terms of renormalized biased operators as 
 \cite{McDonald:2009dh,Chan:2012jj,Assassi:2014fva,Angulo:2015eqa, Desjacques:2016bnm}
\begin{align}\label{eq:G_bias}
\delta_g(\bx)=b_1\delta(\bx) + \frac{b_2}{2} \delta^2(\bx)+ b_{\mathcal{G}_2}\mathcal{G}_2(\bx)\;.
\end{align}
The coefficients $b_1$, $b_2$ and $b_{\mathcal{G}_2}$ are bias parameters whose values depend on the galaxy sample under consideration.
${\mathcal G}_2$ is the tidal Galilean operator defined as:\footnote{This is not to be confused with the second order velocity kernel $G_2$.}
\begin{align}
\mathcal{G}_2(\vx) & \equiv\left[\frac{\partial_i\partial_j}{\nabla^2}\delta(\vx)\right]^2-\delta^2(\vx)\;. 
\end{align}
\vskip 4pt

When galaxies move away or towards the observer with a velocity that differs from the background expansion, their inferred redshift is offset from their true position. These redshift-space distortions can be accounted for by modeling the galaxy velocity field. This has a perturbative component, which follows from solving the equations of motion perturbatively, and a Finger-of-God component, which is caused by very fast galaxies, for example satellite galaxies in virialized halos.  Accounting only for the perturbative velocity component in standard Eulerian perturbation theory, the galaxy density in redshift-space up to second order in the linear density $\delta_1$ becomes 
\begin{align}
    \delta_2(\vx) &= b_1\delta_1(\vx) + f \delta_1^\parallel(\vx) + b_1 F_2[\delta_1,\delta_1](\vx) + \frac{b_2}{2}\delta_1^2(\vx) + b_{\mathcal{G}_2} {\mathcal G}_2[\delta_1,\delta_1](\vx) \notag \\
    &\quad  + f G_2^\parallel[\delta_1,\delta_1](\vx) + b_1f \mathcal{S}_4[\delta_1,\delta_1](\vx)
   + f^2\hat z_i\hat z_j\partial_i\left[ \delta_1^\parallel(\vx) \frac{\partial_j}{\nabla^2}\delta_1(\vx)
    \right]\;.
    \label{eq:delta2}
\end{align}
In this expression, $f$ denotes the logarithmic growth rate, repeated indices are summed over, and $\hat z$ denotes the line-of-sight, which we assume to be along the $z$-axis. 
The operator $F_2$ is defined as
\begin{equation}
  \label{eq:F2Def}
  F_2[a,b](\vk) \equiv \int\frac{\d^3\vq}{(2\pi)^3}  \,\frac{1}{2}\Big[a(\vq)b(\vk-\vq)+b(\vq)a(\vk-\vq)\Big] \, F_2(\vq,\vk-\vq),
\end{equation}
where $F_2(\vk_1,\vk_2)$ is the standard Eulerian perturbation theory kernel for the second order density.
For the second order velocity divergence $G_2$, the tidal field $\mathcal G_2$, and $\mathcal S_4$ (defined in Eq.~\eqref{eq:S4} below), the operators are defined in a similar way. 
We also defined
\begin{align}
    \mathcal O^\parallel &= 
    \hat z_i\hat z_j\frac{\partial_i\partial_j}{\nabla^2}\mathcal O
\end{align}
for an arbitrary field $\mathcal O$. 
In Fourier space, we have
\begin{align}
\label{eq:delta2k}
    \delta_2(\vk) = Z_1(\vk)\delta_1(\vk) + \int_{\vk_1,\vk_2} \delta_D(\vk-\vk_1-\vk_2) Z_2(\vk_1,\vk_2)\delta_1(\vk_1)\delta_1(\vk_2)\;,
\end{align}
where \cite{ScocciRSDbisp98,jeong-phd} 
\begin{align}
  \label{eq:2}
  Z_1(\vk_1) &= b_1 + f\frac{k_{1\parallel}^2}{k_1^2}\;, \\
  Z_2(\vk_1,\vk_2) &=  b_1F_2(\vk_1,\vk_2) + \frac{b_2}{2} + b_{{\mathcal G}_2} {\mathcal G}^2(\vk_1,\vk_2) 
  + f \frac{k_{\parallel}^2}{k^2}  G_2(\vk_1,\vk_2)
  \nonumber \\
&\quad
+ b_1 f \frac{k_{\parallel}}{2}  \left(\frac{k_{1\parallel}}{k_1^2}+\frac{k_{2\parallel}}{k_2^2}\right)
+f^2 \frac{k_{\parallel}}{2}
\left(
\frac{k_{1\parallel}}{k_1^2}\frac{k_{2\parallel}^2}{k_2^2} + \frac{k_{1\parallel}^2}{k_1^2}\frac{k_{2\parallel}}{k_2^2}
\right),
\end{align}
where $\vk=\vk_1+\vk_2$ in the last line and $k_{n\parallel}= \vk_n\cdot \hat z$ is the line-of-sight component of the wavevector $\vk_n$.
\vskip 4pt

\subsection{Bispectrum}

The expression \eq{delta2k} for the galaxy density can be used to compute the galaxy bispectrum in redshift-space at leading order (tree-level) in standard Eulerian perturbation theory.
Considering a single permutation of wavevectors
$\vk_1$, $\vk_2$ and $\vk_3$, the
unsymmetric galaxy-galaxy-galaxy bispectrum $B_\mathrm{ggg}^\text{unsym}$ is
\begin{equation}
  \label{eq:1}
  B_\mathrm{ggg}^\mathrm{unsym}(\vk_1,\vk_2;\vk_3) = 2Z_1(\vk_1)Z_1(\vk_2)Z_2(\vk_1,\vk_2)P_\mathrm{mm}(k_1)P_\mathrm{mm}(k_2),
\end{equation}
where $P_\mathrm{mm}$ is the linear dark matter power spectrum.
The fully symmetric bispectrum  can be obtained with
\begin{equation}
  \label{eq:7}
  B_\mathrm{ggg}^\mathrm{sym}(\vk_1,\vk_2,\vk_3) =
  B_\mathrm{ggg}^\mathrm{unsym}(\vk_1,\vk_2;\vk_3)
+B_\mathrm{ggg}^\mathrm{unsym}(\vk_1,\vk_3;\vk_2)
+B_\mathrm{ggg}^\mathrm{unsym}(\vk_2,\vk_3;\vk_1)\;,
\end{equation}
where $\vk_3=-\vk_1-\vk_2$.
\vskip 4pt

\subsection{Skew-spectra}

Our goal is to estimate the coefficients of all bispectrum contributions, which will be estimators for powers of bias parameters and the logarithmic growth rate $f$ if other cosmological parameters are fixed.\footnote{It is possible to incorporate other cosmological parameters as well, but we leave this for future work.} 
To this end, we first split the bispectrum into contributions $B^{f^n}\propto f^n$ involving different powers of the growth rate $f$,
\begin{equation}
  \label{eq:4}
  B_\mathrm{ggg}^\mathrm{unsym}(\vk_1,\vk_2;\vk_3) =
2P_\mathrm{mm}(k_1)P_\mathrm{mm}(k_2)
 \sum_{n=0}^4 B^{f^n}(\vk_1,\vk_2),
\end{equation}
where we also factored out power spectrum factors for convenience.  We get
\begin{align}
  B^{f^0} =&\;
2 f^0\bigg\{ b_1^3
  F_2(\vk_1,\vk_2)
\label{eq:RSDBhhhf0}
+ b_1^2\frac{b_2}{2} + b_1^2 b_{\mathcal{G}_2} S^2(\vk_1,\vk_2)\bigg\},\\
\nonumber
  B^{f^1} =&\;
f \bigg\{
- b_1^3\frac{k_{3\parallel}}{2} \left(\frac{k_{1\parallel}}{k_1^2}+\frac{k_{2\parallel}}{k_2^2}\right)\\
\nonumber
& \qquad 
+ 
2 b_1^2\left[
F_2(\vk_1,\vk_2) \left(\frac{k_{1\parallel}^2}{k_1^2}+\frac{k_{2\parallel}^2}{k_2^2}\right) + \frac{k_{3\parallel}^2}{k_3^2}G_2(\vk_1,\vk_2)
\right]
\\
&\qquad + 2b_1 \left[\frac{b_2}{2} + b_{\mathcal{G}_2} S^2(\vk_1,\vk_2)\right]\left(\frac{k_{1\parallel}^2}{k_1^2}+\frac{k_{2\parallel}^2}{k_2^2}\right) \bigg\},\\
\nonumber
  B^{f^2} =&\;
f^2 \bigg\{
 -b_1^2\frac{k_{3\parallel}}{2}\left(
\frac{k_{1\parallel}^3}{k_1^4}+\frac{k_{2\parallel}^3}{k_2^4}
+2\frac{k_{1\parallel}}{k_1^2}\frac{k_{2\parallel}^2}{k_2^2} + 
2\frac{k_{1\parallel}^2}{k_1^2}\frac{k_{2\parallel}}{k_2^2} 
\right)
\\
\nonumber
&\qquad
+b_1\left[
\frac{k_{1\parallel}^2}{k_1^2}\frac{k_{2\parallel}^2}{k_2^2}F_2(\vk_1,\vk_2) + \left(\frac{k_{1\parallel}^2}{k_1^2}+\frac{k_{2\parallel}^2}{k_2^2}\right) \frac{k_{3\parallel}^2}{k_3^2}  G_2(\vk_1,\vk_2)
\right]
\\
&\qquad + 2\left[\frac{b_2}{2}+ b_{\mathcal{G}_2} S^2(\vk_1,\vk_2)\right]\frac{k_{1\parallel}^2}{k_1^2}\frac{k_{2\parallel}^2}{k_2^2}
\bigg\},\\
\nonumber
  B^{f^3} =&\;
f^3 \bigg\{
-b_1\frac{k_{3\parallel}}{2}\left[
\frac{k_{1\parallel}^4}{k_1^4}
\frac{k_{2\parallel}}{k_2^2} + \frac{k_{1\parallel}}{k_1^2}\frac{k_{2\parallel}^4}{k_2^4}
+2\frac{k_{1\parallel}^3}{k_1^4}\frac{k_{2\parallel}^2}{k_2^2}
+2\frac{k_{1\parallel}^2}{k_1^2}\frac{k_{2\parallel}^3}{k_2^4}
\right]
\\
&\qquad 
+
\frac{k_{1\parallel}^2}{k_1^2} \frac{k_{2\parallel}^2}{k_2^2} \frac{k_{3\parallel}^2}{k_3^2}
G_2(\vk_1,\vk_2)
\bigg\},
\\
\label{eq:RSDBhhhf4}
  B^{f^4} =&\;
-f^4 
\frac{k_{3\parallel}}{2}\left(
\frac{k_{1\parallel}^3}{k_1^4}\frac{k_{2\parallel}^4}{k_2^4} + \frac{k_{1\parallel}^4}{k_1^4}\frac{k_{2\parallel}^3}{k_2^4}
\right).
\end{align}
Each term depends on a different combination of the growth function
$f$ and bias parameters $b_1$, $b_2$ and $b_{\mathcal G_2}$ (e.g.~the first term depends
on $f^0b_1^3$, the second on $f^0b_1^2b_2$, etc.). 
In total, there are 14 different combinations of parameters.
Our goal is to extract the amplitudes of these $14$ bispectrum contributions to
constrain $f$, $b_1$, $b_2$ and $b_{\mathcal{G}_2}$ (and $\sigma_8^4$ residing in the
overall power spectrum amplitude). This leads to $14$ different
estimators.
\vskip 4pt

To obtain their specific forms, it is instructive to consider the toy
example of estimating the amplitude of a general bispectrum
contribution
\begin{equation}
  \label{eq:8}
  B(\vk_1,\vk_2;\vk_3) = 2P_\mathrm{mm}(k_1)P_\mathrm{mm}(k_2) D(\vk_1, \vk_2)h(\vk_3)
\end{equation}
for general functions $D$ and $h$.   The maximum likelihood estimator for this amplitude involves the cross-spectrum \cite{Schmittfull:2014tca}
\begin{equation}
  \label{eq:9}
  \hat P_{X, Y}(k)=\langle X|Y\rangle(k) \equiv
\frac{1}{4\pi L^3}
\int\d\Omega_{\hat{\vk}} X(\vk) Y(-\vk)
\end{equation}
between the quadratic field
\begin{equation}
  \label{eq:QuadField}
X(\vk) =
\int \frac{\d^3\vq}{(2\pi)^3} D(\vq,\vk-\vq) \delta_R(\vq)\delta_R(\vk-\vq)
\end{equation} 
and the filtered density
\begin{equation}
  \label{eq:10}
  Y(\vk) = h(\vk)\delta(\vk).
\end{equation}
These cross-spectra measure the projection of the observed bispectrum on the theory expectation in a nearly optimal way.\footnote{In presence of noise, all densities should additionally be down-weighted by the noise or filtered by the inverse covariance. }  
To remove the contribution from small-scale modes, we apply Gaussian filters with smoothing scale $R$ to the densities on the right-hand side of \eqq{QuadField}. 
For a tophat filter, this would be equivalent to a $k_\mathrm{max}$ cut in bispectrum analyses, which also makes sure that wavevectors above some threshold are not included.
\vskip 4pt

Following this example, the skew-spectra corresponding to the 14 distinct bispectrum contributions in Eqs.~\eq{RSDBhhhf0}-\eq{RSDBhhhf4} are $\langle \mathcal{S}_n\delta\rangle$, where each quadratic operators $\mathcal S_n$ picks up a different combination of bias parameters and growth rate $f$.
Explicitly, these quadratic operators $\mathcal S_n(\vx)$ are
\begin{align}
    b_1^3: & \qquad \mathcal S_1 = F_2[\delta,\delta]\\
    b_1^2b_2: & \qquad \mathcal S_2 = \delta^2 \\
    b_1^2 b_{\mathcal G_2}: & \qquad \mathcal S_{3} = S^2[\delta,\delta]\\
    b_1^3f: & \qquad \mathcal S_4 = \hat z_i\hat z_j\,\partial_i\left(\delta\frac{\partial_j}{\nabla^2}\delta\right) \label{eq:S4}\\
    b_1^2f: & \qquad \mathcal S_5 = 2F_2[\delta^\parallel,\delta] + G_2^\parallel[\delta,\delta]  \\
    b_1b_2 f: & \qquad  \mathcal S_6 = \delta\delta^\parallel \\
    b_1 b_{\mathcal G_2} f: & \qquad \mathcal S_7 =  S^2[\delta,\delta^{\parallel}] \\
     b_1^2f^2: & \qquad \mathcal S_8 = \hat z_i\hat z_j\partial_i\left(\delta\frac{\partial_j}{\nabla^2}\delta^\parallel
    +2\delta^\parallel \frac{\partial_j}{\nabla^2}\delta
    \right)
    \\
    b_1f^2: & \qquad \mathcal S_9 = F_2[\delta^\parallel,\delta^\parallel] + 2 G_2^\parallel[\delta^\parallel,\delta]\\
    b_2f^2: & \qquad \mathcal S_{10} = \big(\delta^\parallel\big)^2\\
    b_{\mathcal G_2} f^2: & \qquad  \mathcal S_{11} = S^2(\delta^{\parallel}, \delta^{\parallel})\\
    b_1f^3: & \qquad \mathcal S_{12} = \hat z_i\hat z_j\partial_i\left(\delta^{\parallel\parallel}\frac{\partial_j}{\nabla^2}\delta
    +2\delta^\parallel \frac{\partial_j}{\nabla^2}\delta^\parallel
    \right)\\
    f^3: & \qquad \mathcal S_{13} = G_2^\parallel[\delta^\parallel,\delta^\parallel]\\
    f^4: & \qquad \mathcal S_{14} = \hat z_i\hat z_j\partial_i\left(\delta^{\parallel\parallel}\frac{\partial_j}{\nabla^2}\delta^\parallel \right)\;.
\end{align}
In these expressions, all products are in pixel space.
We also defined
\begin{align}
    \mathcal O^{\parallel\parallel} &=
    \hat z_i\hat z_j\hat z_m\hat z_n\frac{\partial_i\partial_j\partial_m\partial_n}{\nabla^4}\mathcal O\;,
\end{align}
as well as operators $\mathcal O[a,b]$ that act on arbitrary fields $a$ and $b$ as in \eqq{F2Def}.
\vskip 4pt

Measuring these 14 skew-spectra derived from the shape of the redshift-space galaxy bispectrum is expected to capture the same information on bias parameters and $f$ as measuring the full bispectrum \cite{MoradinezhadDizgah:2019xun}.
\vskip 4pt

\section{Implementation}

Having derived the skew-spectra that follow from the redshift-space galaxy bispectrum, we proceed by describing our implementation for making analytical predictions for their expectation values and for measuring them from a given galaxy density field.
\vskip 4pt

\subsection{Analytical Predictions}

Analytical predictions for the cross-spectra $\langle \mathcal{S}_n\delta\rangle$ can be obtained by integrating over the theory bispectrum with weights determined by the quadratic operators involved in the skew spectra.
For example, for $\mathcal S_4$,
\begin{align}
P_{\mathcal S_4\delta}(k) = \frac{1}{4 \pi}\int d\Omega_{\hat k} \ k_{\parallel} \int d \Omega_{\hat q} \int \frac{q^2 dq}{(2 \pi)^3} \ \left[\frac{(\vk-\vq)\cdot\hat\vz}{|\vk-\vq|^2} + \frac{\vq\cdot\hat\vz}{q^2}\right]\langle \delta_R(\vq)\delta_R(\vk - \vq)\delta(-\vk)\rangle 
\end{align}
Assuming the line of sight, $\hat \vz$, to be along the $z$-direction, $\vk$ in the $z-y$ plane and $\vq$ to be a general vector, we have
\be
\hat \vz = (0,0,1), \qquad 
\hat \vk =  \left(0,\sqrt{1-\mu_k^2}, \mu_k\right), \qquad 
\hat \vq =  \left(\sqrt{1-\mu_q^2}\cos \phi,\sqrt{1-\mu_q^2}\sin \phi, \mu_q\right).
\ee
The result is a non-separable four-dimensional integral, which we evaluate  numerically using the CUBA library \cite{Hahn:2004fe}. When computing the integrals we restrict the range of  $q$ and $|\vk-\vq|$ to values larger than the fundamental mode of the simulation box to avoid possible divergences.
\vskip 4pt

\subsection{Measurement Pipeline}
To measure the skew-spectra, we use \textsc{nbodykit} \cite{Hand:2017pqn} \href{https://github.com/bccp/nbodykit}{\faGithub} \footnote{\url{https://github.com/bccp/nbodykit}} to compute the quadratic fields $\mathcal{S}_n$ and their cross-spectrum with the density.
To compute the quadratic fields, we Fourier transform two copies of the input density, apply the filters corresponding to each $\mathcal S_n$ by multiplying each copy by appropriate factors in $\vk$, Fourier transform back to real space, and multiply the two fields there.
An accompanying Python software package, \textsc{skewspec}, is available online \href{https://github.com/mschmittfull/skewspec}{\faGithub} \footnote{\url{https://github.com/mschmittfull/skewspec}}.
\vskip 4pt

Measuring all 14 skew-spectra on $512^3$ grids takes about 4 minutes on 14 cores with our implementation.
This could be improved, for example by caching quadratic fields that enter multiple skew spectra, but we expect the current computational cost to be acceptable for most practical purposes.
In terms of scaling, computing skew-spectra scales like $\mathcal{O}(N\log N)$, where $N\simeq 512^3$, which is faster than full bispectrum estimations that scale like $\mathcal{O}(N^2)$.
\vskip 4pt

\section{Numerical Results}\label{sec:sims}

To validate the method we perform numerical tests of the skew-spectra introduced above. A realistic test would be to compare the skew spectra of galaxies in N-body simulations against analytical predictions.
Such a comparison would require a full MCMC analysis to fit the unknown halo/galaxy biases. Since our focus here is to test the validity of the theoretical predictions for skew spectra, we work with data that have a known analytical prediction without requiring any fits of the theory to the measurements. We will perform three such tests in the following, using a synthetic dark matter field generated with perturbation theory, a simulated dark matter field, and a synthetic galaxy density field. These of course are less realistic but more direct and stringent tests of the framework. We defer the full likelihood analysis to future work.
\vskip 4pt

\begin{figure}[htbp!]
\centering
\includegraphics[width=1.0\textwidth]{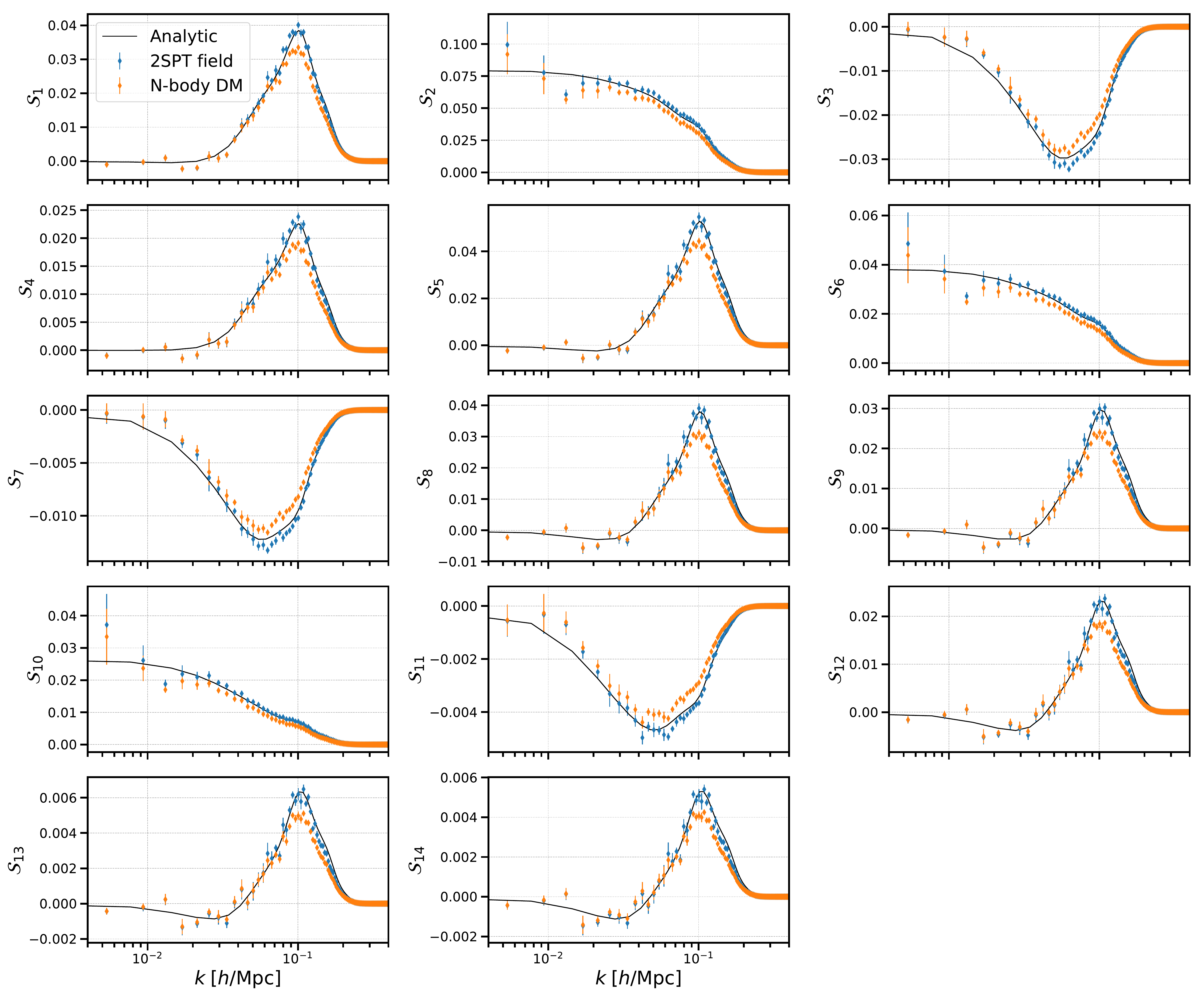}
\caption{The analytical prediction (black lines) matches the skew spectra measured from six realizations of the second order dark matter density field in redshift-space on a $512^3$ grid (blue points). The N-body dark matter density is more nonlinear, which degrades the agreement with theory, especially on small scales (orange points). All curves are normalized to the linear theory prediction for the monopole density power spectrum, i.e.~they show $\langle\mathcal{S}_n[\delta_{2,R}]\delta_2\rangle/P_0(k)$.
The fields entering the quadratic operators $\mathcal{S}_n$ are smoothed with a Gaussian with $R=20 \ \mathrm{Mpc}/h$, while no smoothing is applied to the field entering linearly. This smoothing corresponds to a small-scale cutoff $k_\mathrm{max}$ in the bispectrum and implies that the skew spectra vanish at high $k$.
The plot is for redshift $z=0.6$ and boxsize $L=1500\mathrm{Mpc}/h$. 
}
\label{fig:2SPT}
\end{figure}

\subsection{Comparison with Synthetic 2SPT Field}

As the simplest test of the skew spectrum framework, we generate synthetic 3D dark matter fields in redshift-space using perturbation theory up to second order. To do that, we generate linear density realizations in 3D cubic boxes and compute $\delta_2^\text{RSD}$ using \eqq{delta2} with $b_1=1$, $b_2=0$ and $b_{\mathcal{G}_2}=0$. The resulting skew spectra are shown in \fig{2SPT}.
They match the analytical prediction based on the tree-level SPT bispectrum.
This validates the measurement pipeline and the analytical predictions. Note that small deviations around $k=0.1\;h/\mathrm{Mpc}$ in \fig{2SPT} are due to the 1-loop `222' term (i.e., the correlation of three second-order densities), which contributes when computing skew spectra of $\delta_2$ on the grid, but is not included in the tree-level analytical prediction that only accounts for `211' terms (i.e.~correlations between second-order density and two first-order densities). It is not surprising that 1-loop bispectrum contributions become relevant at $k>0.1\;h/\mathrm{Mpc}$ for tracers like the one we assumed.
When including smaller scales in the quadratic fields $\mathcal S_n$, the tree-level theory prediction tends to deviate more at high $k$ (see Appendix). This is expected because 1-loop terms like the `222' term are increasingly important in this small-scale regime.
\vskip 4pt

\begin{figure}[t]
\centering
\includegraphics[width=1.0\textwidth]{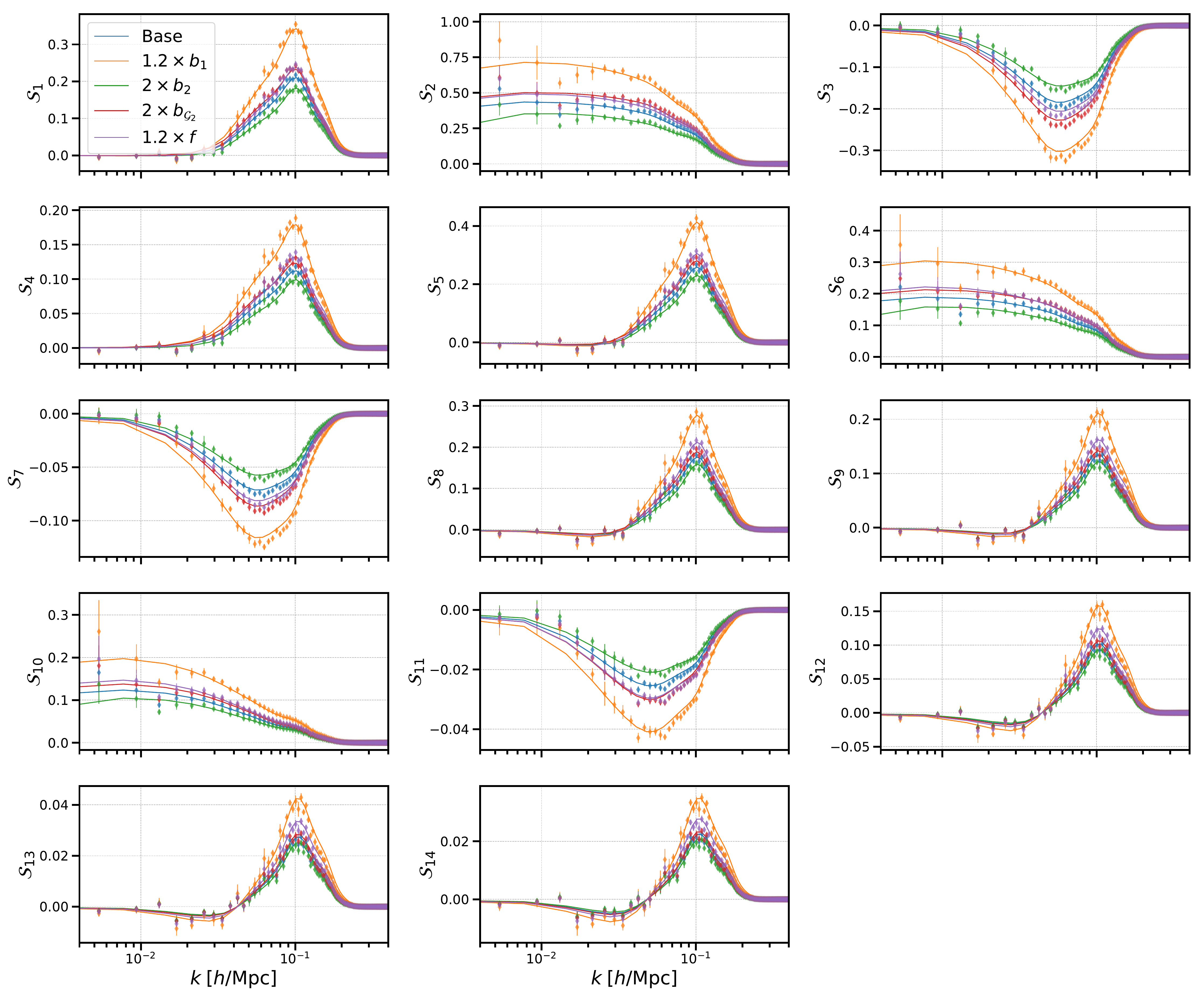}
\caption{Skew spectra for a synthetic galaxy density field with $b_1=2$, $b_2=-0.5$ and $b_{\mathcal{G}_2}=-0.4$ (`Base'). 
Data points show averages over 6 realizations and solid lines are analytical predictions. Different colors show the response of the skew spectra to changing the bias parameters or the logarithmic growth rate $f$.
As in \fig{2SPT}, each curve is normalized by the linear theory DM power spectrum monopole and we apply $R=20\ \mathrm{Mpc}/h$ Gaussian smoothing for the fields entering the quadratic operators.
}
\label{fig:SynGal}
\end{figure}

\subsection{Comparison with N-body Dark Matter Simulations}

To consider larger nonlinearities than those of the perturbative fields above, we run 6 N-body simulations with
\textsc{MP-Gadget} \cite{yu_feng_2018_1451799} \href{https://github.com/MP-Gadget/MP-Gadget}{\faGithub} \footnote{\url{https://github.com/MP-Gadget/MP-Gadget}}, evolving $1536^3$ dark matter particles in a $L=1500\,\mathrm{Mpc}/h$ box to $z=0.6$. 
To account for redshift-space distortions, the DM particles are displaced along the line of sight according to their particle velocity.
We use a random $4\%$ subsample to paint the DM density to a $512^3$ grid and compute its skew-spectra.
The result is shown by the orange points in \fig{2SPT}.
\vskip 4pt

These measurements broadly follow the analytical prediction, but they do not match as well as the 2SPT field considered previously, especially on small scales, $k\gtrsim 0.08\,h/\mathrm{Mpc}$.
This is not surprising because 
the simulated DM density contains strong nonlinearities, both due to nonlinear DM clustering in real space and the nonlinear DM velocity entering the RSD displacements.
Since most galaxies move slower than simulated DM particle velocities, it is reasonable to expect a smaller impact of velocity nonlinearities when working with galaxies rather than DM. 
We do not investigate this further here, noting that one would have to fit bias parameters when working with the galaxy density, which would require the full covariance of the skew spectra, which is beyond the scope of this study.
When including smaller scales in the quadratic fields $\mathcal S_n$, the tree-level theory prediction deviates from the DM measurements at lower $k$ for some of the skew-spectra (see Appendix).
\vskip 4pt

\subsection{Comparison with Synthetic Galaxies}

As a final application, we consider a synthetic galaxy density with perfectly known galaxy bias parameters and no shot noise.
To obtain this, we proceed as for the synthetic DM density above, but set $b_1=2$, $b_2=-0.5$ and $b_{\mathcal{G}_2}=-0.4$ in \eqq{delta2}. 
This roughly represents LRG galaxies observed by spectroscopic surveys like SDSS BOSS or DESI at $z=0.6$, but it is perturbative with less nonlinearity than a fully realistic galaxy density.
The resulting skew spectra are shown in \fig{SynGal}.
The analytical tree-level prediction matches the measured skew spectra well. 
As before, small differences are due to 1-loop `222' correlations that are included in the measurements but not in the analytical prediction.
\vskip 4pt

\fig{SynGal} also shows that the skew spectra change when varying bias parameters or the logarithmic growth rate $f$. 
This can be used to measure combinations of these parameters for real galaxy survey data, where the galaxy bias and logarithmic growth rate are not a priori known. Since the skew spectra probe different parameter combinations than the power spectrum multipoles, combining the measurements breaks parameter degeneracies and is expected to lead to tighter cosmological parameter measurements. In particular, the quadratic biases $b_2$ and $b_{\mathcal{G}_2}$ are difficult to determine from the power spectrum multipoles (assuming realistic survey volume and shot noise), and the skew spectra could be a welcome tool to measure them and improve cosmological power spectrum analyses. We plan to study this in future work, using the covariance matrix between the skew spectra that is needed to obtain parameter constraints.
\vskip 4pt

\section{Conclusions}\label{sec:con}

Previous work studied large-scale structure skew-spectra, i.e.~cross-power spectra between quadratic density fields and the density, as a means to measure the large-scale structure bispectrum in a potentially more convenient manner without losing information \cite{Schmittfull:2014tca,Abidi:2018eyd,MoradinezhadDizgah:2019xun,Dai:2020adm}.
Here, we generalized this approach to redshift-space.
Since all contributions to the tree-level galaxy bispectrum in redshift-space are product-separable in wavevectors, it is possible to extract their amplitudes using skew-spectra.
We found that 14 distinct skew-spectra are required to capture the tree-level galaxy bispectrum information on the growth rate $f$ and the three galaxy bias parameters $b_1$, $b_2$, and $b_{\mathcal{G}_2}$.
This can be used to obtain tighter measurements of these parameters and other cosmological parameters when combined with standard power spectrum analyses.
\vskip 4pt

We implemented a pipeline to measure these skew-spectra, and applied it to a number of different large-scale structure densities in redshift-space.
The measurements agree well with analytical predictions.
We also showed that the dependence of the skew-spectra on $f$ and bias parameters is as expected theoretically.
\vskip 4pt

These results are an important step to measure skew-spectra from galaxy redshift surveys and include them in cosmological large-scale structure data analyses.
The next steps of this program include measuring and modeling skew-spectra for more realistic galaxy densities, including Fingers of God, shot noise, and potentially other systematics, computation of the covariance matrix of the skew-spectra, speed up of theory predictions to enable faster computation of parameter posteriors, inclusion of the survey window function, and inclusion of additional cosmological parameters (which lead to additional skew-spectra). It could also be useful to compress the skew-spectra before running MCMC chains to obtain parameter posteriors \cite{Philcox:2020zyp}. Additionally, it may be interesting to consider 1-loop corrections to the bispectrum, incorporate 4-point information by correlating cubic fields with the density \cite{Abidi:2018eyd}, or use skew-spectra to study new halo biases \cite{Fujita:2020xtd}. With this, skew-spectra can become a valuable tool to incorporate 3-point information in large-scale structure analyses, complementing other approaches that measure the bispectrum or 3-point correlation function more directly. 
\vskip 4pt

\subsection*{Acknowledgements}
It is our pleasure to thank M.~Ivanov, M.~Simonovi\'c, O.~Philcox and M.~Zaldarriaga for helpful discussions, as well as M.~Abidi and Z.~Vlah for comments on the draft. M.S.~acknowledges support from the Corning Glass Works Fellowship and the National Science Foundation. A.M.D.~is supported by the SNSF project ``The  Non-Gaussian  Universe  and  Cosmological Symmetries", project number:200020-178787. 
\vskip 4pt

\appendix

\section{Including Smaller Scales in Quadratic Fields}
\begin{figure}[htbp!]
\centering
\includegraphics[width=1.0\textwidth]{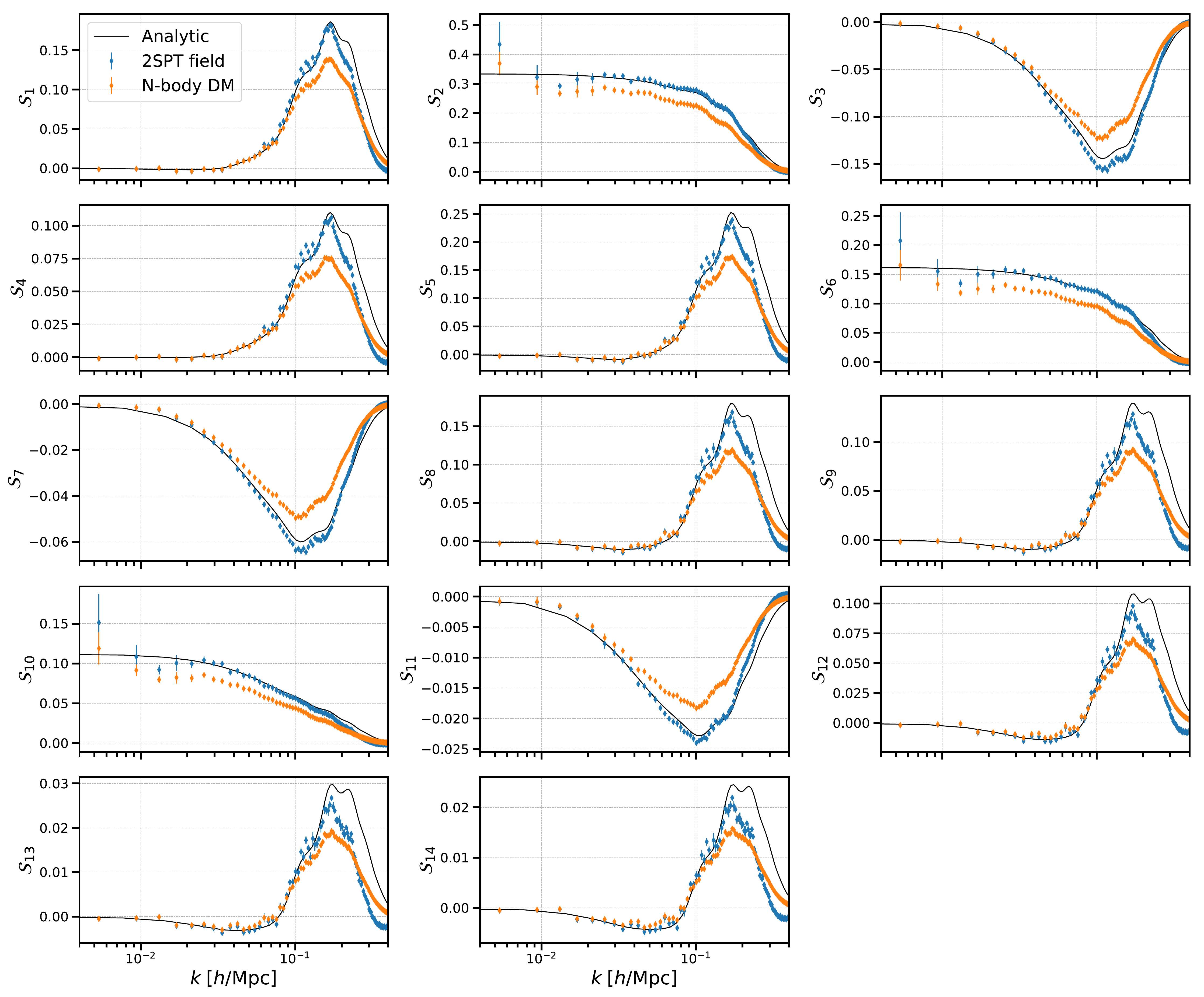}
\caption{Same as \fig{2SPT} but using $R=10\,h^{-1}\mathrm{Mpc}$ Gaussian smoothing for the fields entering quadratic operators $\mathcal S_n$.
}
\label{fig:2SPT10}\vspace{-.2in}
\end{figure}

\begin{figure}[tbp]
\centering
\includegraphics[width=1.0\textwidth]{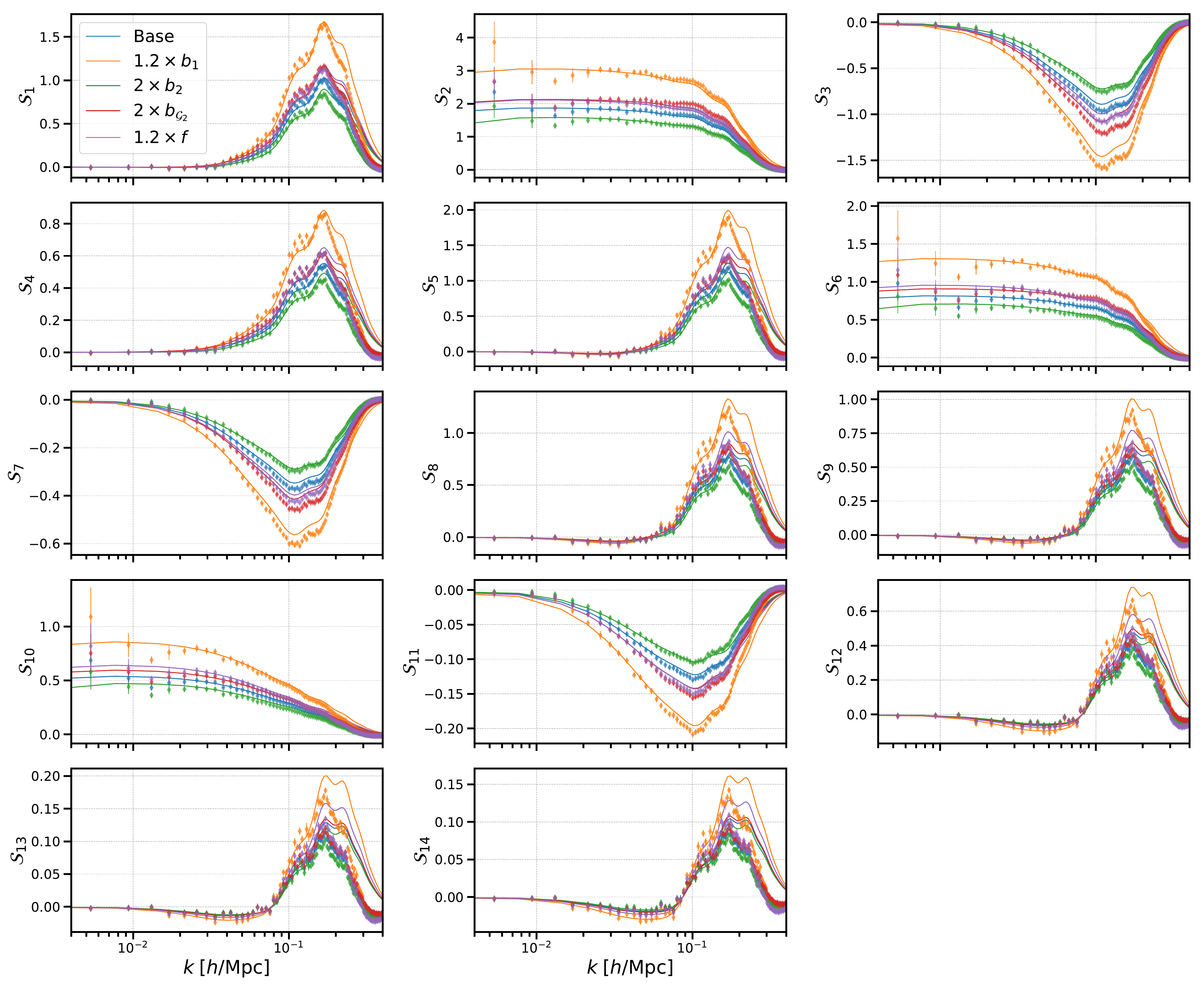}
\caption{Same as \fig{SynGal} but using $R=10\,h^{-1}\mathrm{Mpc}$ Gaussian smoothing for the fields entering quadratic operators $\mathcal S_n$.
}
\label{fig:SynGal10}
\end{figure}

In this appendix we show results when including smaller scales in quadratic fields. For that, we apply $R=10\,h^{-1}\mathrm{Mpc}$ Gaussian smoothing for the fields entering the quadratic operators $\mathcal S_n$, instead of $R=20\,h^{-1}\mathrm{Mpc}$, which is used in the main text. The results are shown in Figs.~\ref{fig:2SPT10} and \ref{fig:SynGal10}. 
The amplitudes of the skew-spectra and their signal-to-noise ratios are substantially larger, which is expected because more modes are included. However, this comes at the expense of larger deviations from the tree-level bispectrum theory prediction, especially at high $k$.In that regime, including 1-loop corrections to the bispectrum model could be very helpful.

For some skew-spectra, especially $\mathcal S_2, \mathcal S_6$ and $\mathcal S_{10}$, the theory does not match the N-body DM measurements even at very low $k$, even though it does match for the synthetic 2SPT DM field and the synthetic galaxy density. This suggests that the low-$k$ amplitude of these three skew-spectra is rather UV-sensitive when using $R=10\,h^{-1}\mathrm{Mpc}$ smoothing. 
This deserves further investigation, as it is not clear from our numerical experiments to what extent this will be relevant for more realistic tracers.
\vskip 4pt

\bibliographystyle{utphys}
\bibliography{refs}

\end{document}